\documentclass[useAMS,usegraphicx]{mn2e}
\usepackage{times}
\usepackage{rotate}
\usepackage{amssymb}
\usepackage{multirow}
\title[Determining the sensitivity of X-ray imaging
  observations and the X-ray number counts]{A new method for determining the sensitivity of X-ray
  imaging observations and the X-ray number counts}  
\author[Georgakakis et al. ] {A. Georgakakis\thanks{Marie Curie
    fellow}, K. Nandra, E. S. Laird, J. Aird, M. Trichas
  \\ \\
  Astrophysics Group, Blackett Laboratory, Imperial College, Prince
  Consort Rd , London SW7 2AZ, UK\\
}
\begin{document}
\maketitle

\begin{abstract}  We   present  a  new  method   for  determining  the
sensitivity  of X-ray imaging  observations, which  correctly accounts
for the observational biases  that affect the probability of detecting
a source  of a given X-ray flux,  without the need to  perform a large
number of  time consuming  simulations. We use  this new  technique to
estimate the  X-ray source counts in different  spectral bands (0.5-2,
0.5-10, 2-10 and 5-10\,keV)  by combining deep pencil-beam and shallow
wide-area {\it Chandra} observations.  The  sample has a total of 6295
unique sources over an area of $\rm 11.8deg^2$ and is the largest used
to date to  determine the X-ray number counts.   We determine, for the
first time,  the break flux  in the 5-10\,keV  band, in the case  of a
double power-law source count distribution.  We also find an upturn in
the 0.5-2\,keV  counts at fluxes  below about $\rm  6\times10^{-17} \,
erg \, s^{-1} \, cm^{-2}$.  We  show that this can be explained by the
emergence  of normal  star-forming galaxies  which dominate  the X-ray
population  at  faint  fluxes.   The  fraction of  the  diffuse  X-ray
background resolved into point  sources at different spectral bands is
also estimated. It is argued that a single population of Compton thick
AGN cannot  be responsible for the entire  unresolved X-ray background
in the energy range 2-10\,keV.
\end{abstract}
\begin{keywords}  
  methods: data analysis -- methods: miscellaneous -- methods:
  statistical -- surveys -- X-rays: galaxies -- X-rays: diffuse
  background  
\end{keywords} 

\section{Introduction}\label{sec_intro}

X-ray  observations  have complex  instrumental  effects  that have  a
strong impact on the detection  probability of point sources. The size
and  the shape of  the Point  Spread Function  (PSF) for  example vary
across the detector.   Also, the mirrors of X-ray  telescopes are more
efficient at  collecting photons from  sources close to the  centre of
the field of view  (vignetting).  This loss of sensitivity effectively
translates  to  a  reduction  of  the exposure  time  with  increasing
off-axis angle.   In addition to  the instrumental effects  above, the
application of  any source detection  software on an X-ray  image also
introduces biases (e.g.  Kenter  \& Murray 2003; Wang 2004).  Brighter
sources  have a higher  probability of  detection compared  to fainter
ones.  Background fluctuations result  in spurious detections that are
inevitably  present,   hopefully  in  small  numbers,   in  any  X-ray
catalogue. Statistical  variations of the source  counts combined with
the  steep $\log  N  - \log  S$  of the  X-ray  population result  in
brighter measured  fluxes for the  detected sources compared  to their
intrinsic  ones (Eddington  bias).   This effect  becomes more  severe
close to the detection threshold of a given X-ray observation.

For a  wide range  of applications it  is important to  quantify these
effects accurately in order to  understand the type of sources a given
X-ray observation is (or is not) sensitive to.  For example, the large
scale structure of X-ray sources  is often estimated using the angular
or the  spatial correlation functions  (e.g.  Basilakos et  al.  2004,
2005; Gilli et  al. 2005; Miyaji et al. 2007).   For this exercise one
needs  to construct  a  simulated comparison  sample  of sources  with
random  spatial  distribution across  the  surveyed  area. These  mock
catalogues should  follow the  same instrumental and  source detection
related biases as the real sample. If not any recovered signal will be
heavily  contaminated.  Also,  for  the estimation  of the  luminosity
function  of X-ray  selected populations  one needs  to  determine the
volume of the survey which is accessible to a source with a particular
flux  and spectral  shape  (e.g.   Hasinger et  al.   2005; Nandra  et
al. 2005a; Aird et al. 2008). Understanding the sensitivity of a given
X-ray observation to AGN of  variable obscuration and luminosity is of
key importance for  studies of the diffuse X-ray  background (e.g.  La
Franca 2005; Akylas  et al.  2006; Gilli, Comastri  \& Hasinger 2007).
Finally, in order to constrain  the number density of X-ray sources to
the faintest fluxes accessible to  a given observation it is essential
to have an  accurate estimate of the total surveyed  area over which a
source of a given flux can be detected (e.g.  Kim et al. 2007a).

The point source  selection function of a given  X-ray observation can
be represented in the form  of a sensitivity map, which should provide
an estimate  of the probability that a  source with a flux  $f_X$ in a
certain  energy  band  will  be  detected across  the  detector.   One
approach to construct  such a map is to perform a  large number of ray
tracing  simulations  (e.g.   Cappelluti  et  al.  2007;  Kim  et  al.
2007a). Artificial sources  with a wide range of  fluxes are placed at
different pixels  on the detector  assuming a realistic model  for the
(instrumental and  cosmic) background. The fraction  of sources picked
up by the detection method as a function of detector position and flux
provides   an  estimate   of   the  sensitivity   of  the   particular
observation. This approach however, is time consuming and difficult to
apply  to large numbers  of observations  with different  setups. These
simulations also cannot fully correct for the Eddington bias, at least
not for individual sources.   This problem has been addressed recently
by  Kenter \&  Murray  (2003) who  presented  a novel  method for  the
construction  of  X-ray  number  counts  that accounts  for  both  the
Eddington bias of individual detections  as well as the variable point
source detection threshold across the survey area. Wang (2004) further
developed this technique to avoid  the need for cumbersome ray tracing
simulations.

In  this paper  we  extend  these previous  studies  by presenting  an
improved  method for  constructing point  source sensitivity  maps for
X-ray imaging observations.  The backbone of our approach is the point
source detection method presented  by Nandra et al.  (2005b). Combined
together these  two techniques provide  a simple and efficient  way of
detecting sources and  determining accurately their selection function
self  consistently by  taking  into account  both instrument  specific
effects and  the source detection  biases.  As a demonstration  of our
method we present  the X-ray number counts in  different energy bands,
0.5--2,  0.5--10, 2--10 and  5--10\,keV.  Future  applications include
the determination of the angular correlation function of X-ray sources
and the  estimation of the  luminosity function of AGN.   Finally, our
method is build  around {\it Chandra} data but  can be easily extended
to {\it XMM-Newton}.

\section{Source detection}\label{sec_det}

The  point  source  detection  method   has  a  central  role  in  the
construction  of  sensitivity maps.   The  adopted  approach is  fully
described by Nandra et al.   (2005b) and has recently been extended to
use a  new set of  PSFs generated by  MARX (Model of AXAF  Response to
X-rays)  as discussed  in Laird  et al.   (2008 in  prep.). Elliptical
apertures are used to estimate  the Encircled Energy Fraction (EEF) as
a  function  of  semi-major  axis  radius.  However,  for  the  source
detection and  sensitivity map construction we  use circular apertures
with  areas equal  to  that of  the  70 per  cent  EEF ellipses.   The
uncertainty  introduced by this  approximation is  $<2$ per  cent. The
source extraction  is based on pre-selection  of positive fluctuations
using the {\sc wavdetect} task  of CIAO at a low probability threshold
of  $10^{-4}$.   The  total  counts  (source and  background)  at  the
position of each candidate source  are then extracted using a circular
area  equal to  the 70  per cent  EEF elliptical  aperture.   The mean
expected background within the detection cell is determined by scaling
the  counts from  a local  annulus centred  on the  source  with inner
aperture equal  to 1.5 times the 90  per cent EEF radius  and width of
50\,arcsec.   The probability that  the candidate  source is  a random
fluctuation   of  the   background  is   estimated   assuming  Poisson
statistics.  A threshold of  $<4\times10^{-6}$ is adopted for a source
candidate to be considered a detection.  At this level about 0.5 false
sources per  {\it Chandra} image are  expected (Laird et  al.  2008 in
prep.).   This method  is  simple and  efficient  resulting in  higher
sensitivity  and  fewer spurious  detections  compared to  alternative
wavelet-based only techniques.  However, the most important feature of
the  method is  that  the detection  cell  has a  fixed  size at  each
position on the detectors,  making the construction of the sensitivity
maps straightforward.

\section{Simulations}\label{sec_sim}

We use MARX to simulate {\it Chandra} observations and to validate the
method for constructing sensitivity maps.  MARX is a suite of programs
that perform  detailed ray tracing  simulations to determine  how {\it
Chandra} responds to astrophysical  sources. In this study we simulate
point  sources  in  the   2-10\,keV  spectral  band  with  a  built-in
distribution per  unit X-ray flux interval, $dN/df_X$,  that follows a
double power-law of the form

\begin{equation}\label{eq_dnds}  
\frac{dN}{df_X} =  \left\{\begin{array}{ll}  K \,
\bigl( \frac{f_X}{f_{ref}} \bigr) ^{\beta_1}, & f_X<f_b,  \\ 
 & \\ 
K^\prime \,  \bigl( \frac{f_X}{f_{ref}} \bigr) ^{\beta_2}, & f_X \ge f_b, \\
\end{array} \right.
\end{equation}

\noindent where the normalisation  constants $K$ and $K^\prime$ follow
the relation $K^\prime = K\,(f_b/f_{ref})^{\beta_1-\beta_2}$, $f_b$ is
the  X-ray flux  of the  break of  the double  power-law, $f_{ref}=\rm
10^{-14}\, erg  \, s^{-1} \,  cm^{-2}$ is the normalisation  flux and
$\beta_1$, $\beta_2$ are the  power-law indices for fluxes fainter and
brighter  than the  break  flux respectively.  The cumulative counts 
follow the relations

\begin{equation}\label{eq_cum} 
N(>f_X) =  \left\{\begin{array}{ll}

K \, \frac{f_{ref}}{1+\beta_1} \Bigl( \bigl( \frac{f_b}{f_{ref}} \bigr) ^{1+\beta_1} - \bigl( \frac{f_X}{f_{ref}} \bigr) ^{1+\beta_1} \Bigr) +  & \\ 
K^\prime \,  \frac{f_{ref}}{1+\beta_2} \bigl( \frac{f_b}{f_{ref}} \bigr) ^{1+\beta_2},  & f_X<f_b, \\ 
 & \\
K^\prime \,  \frac{f_{ref}}{1+\beta_2} \bigl( \frac{f_X}{f_{ref}} \bigr) ^{1+\beta_2}, & f_X \ge f_b, \\
\end{array} \right.
\end{equation}

\noindent  For  the  simulations  we  adopt  the  best-fit  parameters
determined  by   Kim  et  al.    (2007a)  for  the   2-8\,keV  counts:
$\beta_1=-1.58$,          $\beta_1=-2.59$         and         $f_b(\rm
2-10\,keV)=2\times10^{-14} \,  erg \,  s^{-1} \, cm^{-2}$.   The break
flux determined by Kim et al.  (2007a) in the 2-8\,keV band is shifted
to   the   2-10\,keV   energy   range  assuming   $\Gamma=1.4$.    The
normalisation  is fixed  so that  there  are 7000  sources per  square
degree brighter than $f_X(\rm 2 - 10 \, keV) = 10^{-16} \, erg \, s \,
cm^{-2}$, i.e.  similar to the observed density of X-ray sources.

We adopt an X-ray spectrum with $\Gamma=1.4$ and a total exposure time
of 200\,ks.  Simulated ACIS-I event files are  constructed by randomly
placing  within the  {\it Chandra}  field of  view point  sources with
fluxes  in the  range $f_X(\rm  2  - 10\,  keV )  = 5\times10^{-17}  -
10^{-12}\,  erg \,  s^{-1} \,  cm^{-2}$.  MARX  does not  simulate the
Chandra background.  This  is added to the simulated  images using the
quiescent background event files produced by the ACIS calibration team
using blank sky observations. In  the next sections it is demonstrated
that  using the  proposed method  for quantifying  the  sensitivity of
X-ray  imaging  observations we  can  successfully  recover the  input
source count distribution.  

\begin{figure}
\begin{center}
 \rotatebox{0}{\includegraphics[height=0.9\columnwidth]{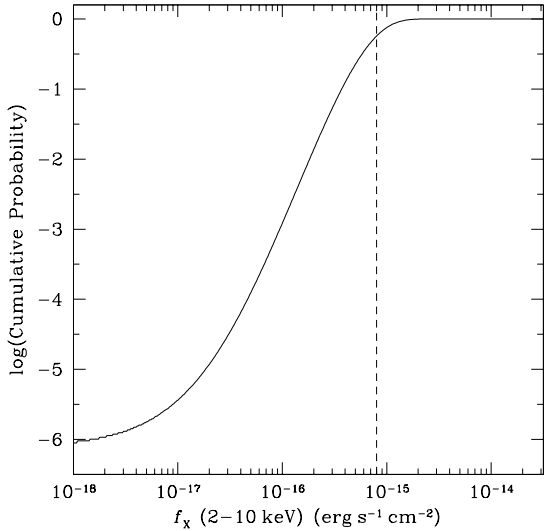}}
\end{center}
\caption{ Probability that a source with  hard flux $f_X(\rm 2 - 10 \,
 keV)$ and  a power-law spectrum with $\Gamma=1.4$  is detected within
 an  element where  the  mean  local background  is  $B=0.14$ and  the
 detection threshold  is $L=5$,  typical of the  aimpoint region  of a
 200\,ks  Chandra  observations. The  vertical  dashed  line shows  the
 approximate  limiting flux  of such  an observation  (e.g.  Nandra et
 al. 2005b). Fluctuations of  the total counts (source and background)
 within the  cell have a  finite probability of producing  a detection
 even if the real source has  flux well below the formal survey limit.
 }\label{fig_element}
\end{figure}

\begin{figure}
\begin{center}
 \rotatebox{0}{\includegraphics[height=0.9\columnwidth]{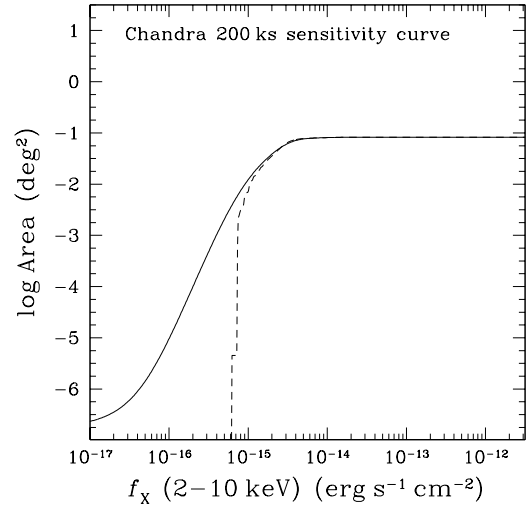}}
\end{center}
\caption{
 Sensitivity curve in the 2-10\,keV spectral band of a simulated 200\,ks
 {\it Chandra} observations estimated using the method
 described in this paper (continuous curve). The dashed line is the
 area curve that does not account for the Poisson probability density
 distribution of the total observed counts in the detection cell. 
 }\label{fig_acurve}
\end{figure}

\section{Sensitivity map construction}\label{sec_sense}

The sensitivity  map is an estimate  of the probability  that a source
with flux $f_X$  in a certain energy band will  be detected across the
detector.  This  probability depends on  the detector characteristics,
the  adopted observational strategy  and the  specifics of  the source
detection. Instead of extensive  simulations one can use statistics to
approximate  the source detection  process, as  this is  essentially a
Poisson experiment. Any source extraction method estimates the Poisson
probability that  the observed counts  in a detection cell  arise from
random fluctuations  of the  background.  The important  parameters in
this exercise are  the size and shape of the  detection cell, which is
well  defined  in  our   source  detection  method,  and  the  Poisson
probability threshold,  $P_{thresh}$, below which an  excess of counts
is considered  a source. By  fixing $P_{thresh}$ one sets  the minimum
number of photons in a cell, $L$, for a formal detection.

The observed counts in a detection cell have a background and possibly
a source component. Either of them  can fluctuate so that their sum is
higher than $L$. This either produces a spurious source in the case of
a  background fluctuation or  can make  a faint  source appear  with a
brighter observed  flux, i.e. the  Eddington bias. Both  these effects
are part of  the source selection process and  should be accounted for
by the sensitivity map.

The  first  step  to  construct  such  an image  is  to  estimate  the
source-free background across the  detector.  This has an instrumental
and a cosmic component, with  the latter coming from sources below the
sensitivity of the observation or photons in the extended wings of the
PSF of detected sources. The  background map is estimated using custom
routines  to first  remove  the  counts in  the  vicinity of  detected
sources using  an aperture size  that is 1.5  times larger the  90 per
cent EEF  radius. Pixel values in  the source regions  are replaced by
sampling  from the distribution  of pixel  values in  local background
regions. These are defined by annuli centred on each source with inner
apertures equal to 1.5 times the  90 per cent EEF radius and widths of
50\,arcsec.  The resulting maps can  then be used to estimate the mean
expected  background  counts  within  any detection  cell,  $B$.   The
cumulative  probability  that  the  observed counts  in  a  particular
detection cell will exceed $L$ is

\begin{equation}\label{equ_poisson}  
P_{B}(\ge L) = \gamma(L, B),
\end{equation} 

\noindent 
where the function $\gamma(a,x)$ is defined

\begin{equation}  
\gamma(a, x) = \frac{1}{\Gamma(a)} \int_{0}^{x} e^{-t} \, t^{a-1}
\, dt.
\end{equation} 

\noindent Adopting  a detection threshold $P_{thersh}$  one can invert
equation  \ref{equ_poisson}  numerically  to  estimate  the  (integer)
detection  limit  $L$  for   a  cell  with  mean  expected  background
$B$. Repeating this exercise for  different cells across the image one
can  determine   $L$  as   a  function  of   position  (x,y)   on  the
detector. This 2-D  image of $L$ values is  the sensitivity map.  Note
that the sensitivity map is  independent of the  spectral shape of
the source.   A useful 1-D representation  of this image,  with a wide
range of  applications, is the total  detector area in  which a source
with flux $f_X$  can be detected.  The cumulative  distribution of the
area  plotted  as  function  of  $f_X$  is  often  referred  to  as  a
sensitivity curve. This is constructed  as follows.  For a source with
flux $f_X$ and a given  spectral shape ($\Gamma=1.4$ in this paper) we
can  determine  the probability  of  detection  in  a cell  with  mean
background $B$ and detection limit  $L$.  The total observed counts in
the  cell  are  $T=B+S$,  where   $S$  is  the  mean  expected  source
contribution.  In  practise this  depends on the  observation exposure
time,  the  vignetting of  the  field at  the  position  of the  cell,
embodied in  the exposure  map, and the  fraction of the  total source
counts in the cell because of the PSF size.

\begin{equation}\label{equ_source}
S = f_X \times t_{exp} \times C \times \eta.
\end{equation} 

\noindent  Where  $t_{exp}$  is  the  exposure time  at  a  particular
position  after  accounting  for  instrumental  effects,  $C$  is  the
conversion  factor from  flux to  count  rate for  the adopted  source
spectral  shape and  $\eta$ is  the encircled  energy fraction  at the
particular position.   Both $B$ and $S$ fluctuate  and therefore using
equation  \ref{equ_poisson}  the  probability  their sum  exceeds  the
detection  threshold is  $P_{T,f_X}(\ge  L) =  \gamma(L, T)$.   Figure
\ref{fig_element}  plots  $P_{T,f_X}(\ge   L)$  against  $f_X$  for  a
particular choice of  $B$ and $L$ values, typical  of a detection cell
close to  the aimpoint of a  200\,ks {\it Chandra}  observation in the
2-10\,keV  energy  band.   At  faint  fluxes, $S  \rightarrow  0$  and
$P_{T,f_X}(\ge L) \rightarrow P_{thresh}$,  because this is the finite
probability of  a random background  fluctuation above the  limit $L$,
i.e.  spurious detections.

The  sensitivity   curve  is  the   sum  of  the   $P_{T,f_X}(\ge  L)$
distributions  of  individual detection  cells.   This is  graphically
shown in Figure \ref{fig_acurve} for the 2-10\,keV band of a simulated
200\,ks Chandra pointing.  For comparison we also show the sensitivity
curve  of this  field  calculated adopting  the  standard approach  of
assigning  a  single limiting  flux  to  a  detection cell.   This  is
estimated by assuming  that the minimum net counts for  a source to be
detected in the  cell are $S=L-B$.  This method  ignores the fact that
the total observed photons in a cell, source and background, fluctuate
according  to  the Poisson  distribution.   These fluctuations  become
increasingly important toward the low count limit, i.e.  faint fluxes
and/or low  background.  The method presented  here correctly accounts
for this effect.  Also, the  finite number of spurious detections that
are inevitably  present in any  source catalogue is factored  into the
sensitivity curve estimation.

\begin{figure}
\begin{center}
 \rotatebox{0}{\includegraphics[height=0.9\columnwidth]{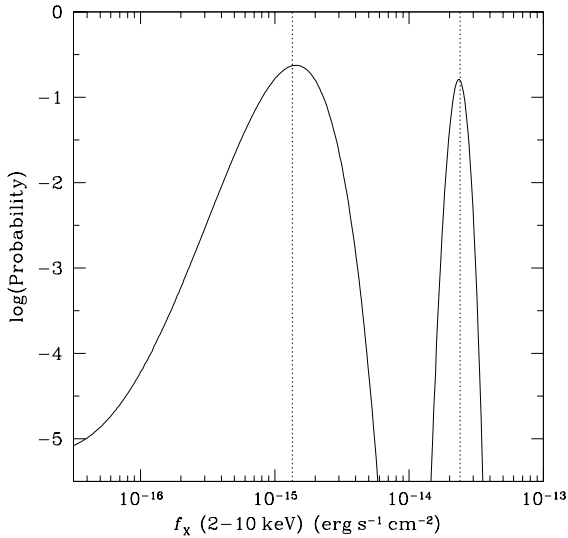}}
\end{center}
\caption{  Flux   probability  density  distribution   estimated  from
equation  \ref{eq_source} for  sources  detected in  the 200\,ks  {\it
Chandra} ACIS-I  simulation.  The  curves estimate the  probability of
observing  exactly $N$ counts  (source and  background) from  a source
with  flux $f_X$.  The  normalisation is  arbitrary.  Examples  of two
simulated  sources are shown.   The faint  one has  a total  number of
counts  within the 70  per cent  EEF cell  of $T=9$  and a  mean local
background  of $B=0.41$  counts.  For  the bright  source  $T=120$ and
$B=3.06$ counts.  The  dotted vertical lines mark the  input fluxes of
the two sources  in the simulated field, $f_X(\rm  2-10) = 1.35 \times
10^{-15}$  and $2.40  \times 10^{-14}  \,  erg \,  s^{-1} \,  cm^{-2}$
respectively. The contribution of each source to the number counts is 
estimated  by dividing  the probability  distribution curves  with the
sensitivity curve of Fig. \ref{fig_acurve}.  
}\label{fig_source}
\end{figure}

\begin{figure}
\begin{center}
 \rotatebox{0}{\includegraphics[height=0.9\columnwidth]{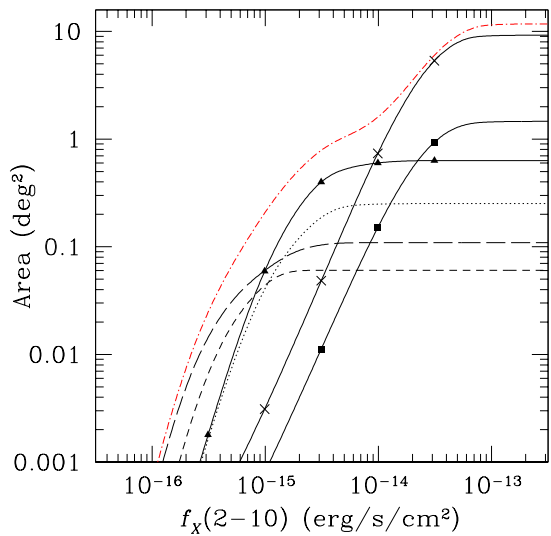}}
\end{center}
\caption{  Sensitivity curve  in  the hard  band  (2-10\,keV) for  the
different  {\it Chandra}  surveys used  to determine  the differential
X-ray  source  counts.  XBOOTES:  continuous line  with  the  crosses;
ELAIS-N1:  continuous line  marked with  squares; EGS:  The continuous
line marked  with triangles; ECDF-S: dotted  line; CDF-S: short-dashed
line; CDF-N: longed-dashed line.  The (red) dashed-dotted line line is
the sum of the individual curves. The different slopes between the
shallow and the deep surveys is the result of the outer regions
becoming background limited. }\label{fig_area} 
\end{figure}

\section{X-ray number count estimation}\label{sec_counts}

In  this  section we  describe  the  methodology  for determining  the
surface density  of X-ray  sources on  the sky as  a function  of flux
using  the  new  approach  for  constructing  sensitivity  maps.   The
standard method  of estimating number counts,  cumulative $N(>f_X)$ or
differential $dN/df_X$, is to weigh each source of flux $f_X$ with the
inverse of the solid angle in  which it can be detected.  The drawback
of this  approach is  that it assigns  a single  flux to a  source and
therefore does not account for the fact that the observed total counts
in the detection cell can be produced by sources with a range of $f_X$
according to  a probability distribution  that can be  estimated using
Poisson statistics. If the mean expected total counts in the detection
cell are $T=S+B$,  where $S$ is determined as a  function of flux from
equation  \ref{equ_source}  and  $B$  is  the  local  background  value
then  the probability of finding the observed number of total counts
$N$ (source and background) is 

\begin{equation}\label{eq_source} 
P(f_X, N) = \frac{T^N \, e^{-T}}{N!} \, f_X^{\beta}.  
\end{equation} 

\noindent where the last term,  $f_X^\beta$, is for the Eddington bias
and  assumes  that  the  differential  X-ray source  counts  follow  a
power-law of  the form ${\rm dN/d}f_X  \propto f_X^{\beta}$.  Equation
\ref{eq_source} is graphically shown in Figure \ref{fig_source} in the
case of  a particular source  drawn from the simulations  described in
section  \ref{sec_sim}.  The  X-ray  number counts  are determined  by
simply dividing the sum of the probability distributions of individual
sources,  $P(f_X, N)$, with  the sensitivity  curve determined  in the
previous section.   This approach has  the advantage that  it accounts
for  source  and  background  fluctuations,  the  Eddington  bias  and
spurious detections in the catalogue.

Determination of $P(f_X, N)$  for individual sources requires however,
knowledge  of  the source  count  power-law  index  $\beta$.  This  is
estimated by applying maximum  likelihood (ML) methods to the unbinned
data.  The probability of  the source $i$ with total number of counts
$N_i$ in the surveyed area is  

\begin{equation}  
P_i=\frac{\int{P(f_X,N_i) \,df_X}}{\int{dN/df_X \, A(f_X) \, df_X}}.
\end{equation} 

\noindent The likelihood  of a particular set of  data is estimated by
multiplying  the probabilities  $P_i$ of  individual sources.   We can
then  estimate   the  power-law  index  $\beta$   that  maximises  the
likelihood. It is straightforward to  generalise the form for $P_i$ in
the case of a source count distribution that follows the double
power-law of equation \ref{eq_dnds}. 

We  demonstrate  this technique  using  the  simulations described  in
section \ref{sec_sim}.  The source detection code described in section
\ref{sec_det}  and  the sensitivity  map  construction  method of  the
previous section are applied to a  total of 10 mock Chandra fields. We
recover the input  source count distribution in the  2-10\,keV band by
combining  the 10  individual simulated  source lists.   Using  the ML
method we estimate the bright  and faint-end power-law indices as well
as  the  break flux,  $\beta_1  =  -1.62^{+0.08}_{-0.06}$, $\beta_2  =
-2.73^{+0.30}_{-0.37}$ and $f_b=\rm (2.0\pm0.6)\times10^{-14}\, erg \,
s^{-1} \, cm^{-2}$, in good  agreement with the input values listed in
section \ref{sec_sim}.

\begin{figure*}
\begin{center}
 \rotatebox{0}{
  \includegraphics[height=0.9\columnwidth]{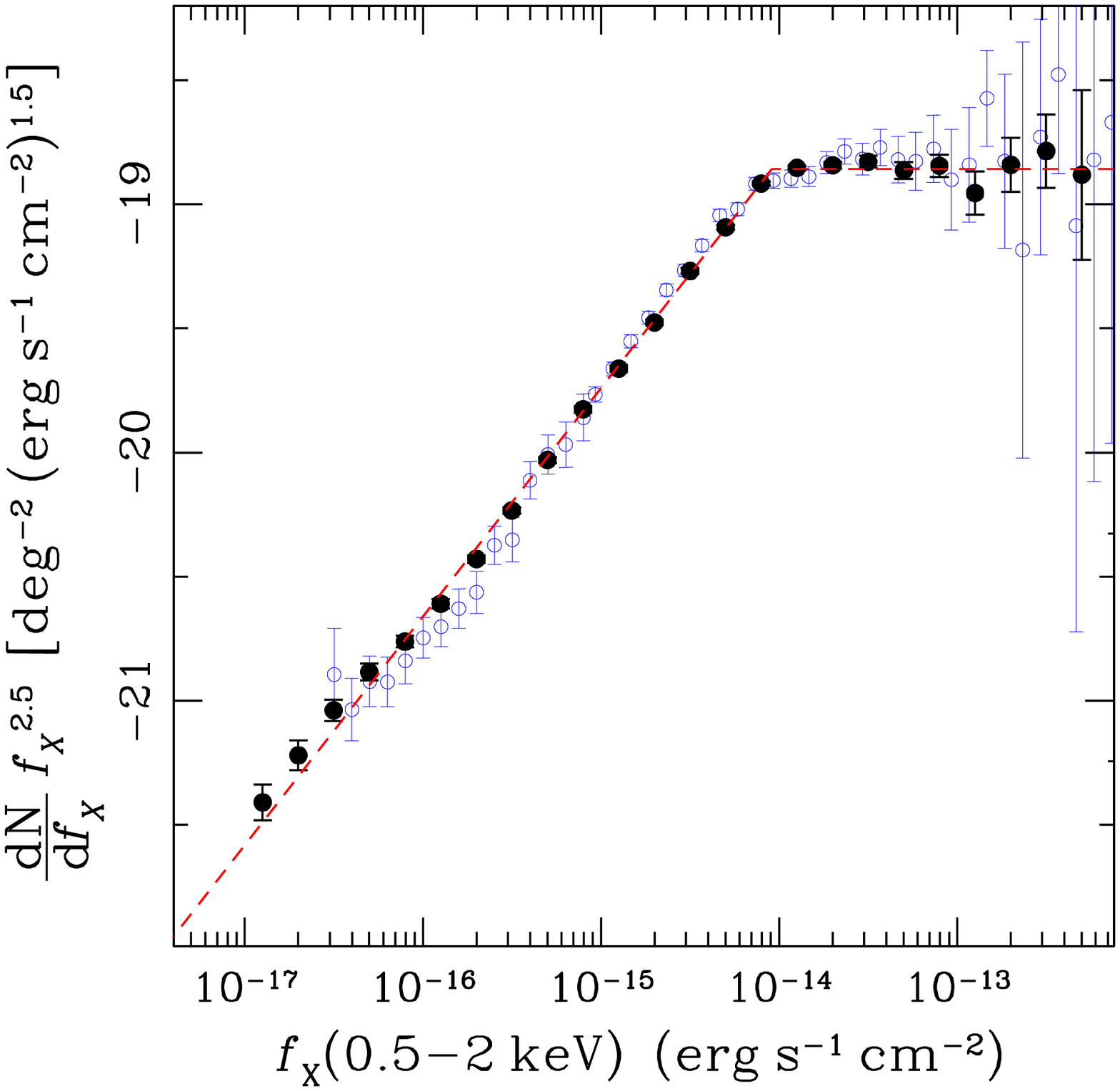}
  \includegraphics[height=0.9\columnwidth]{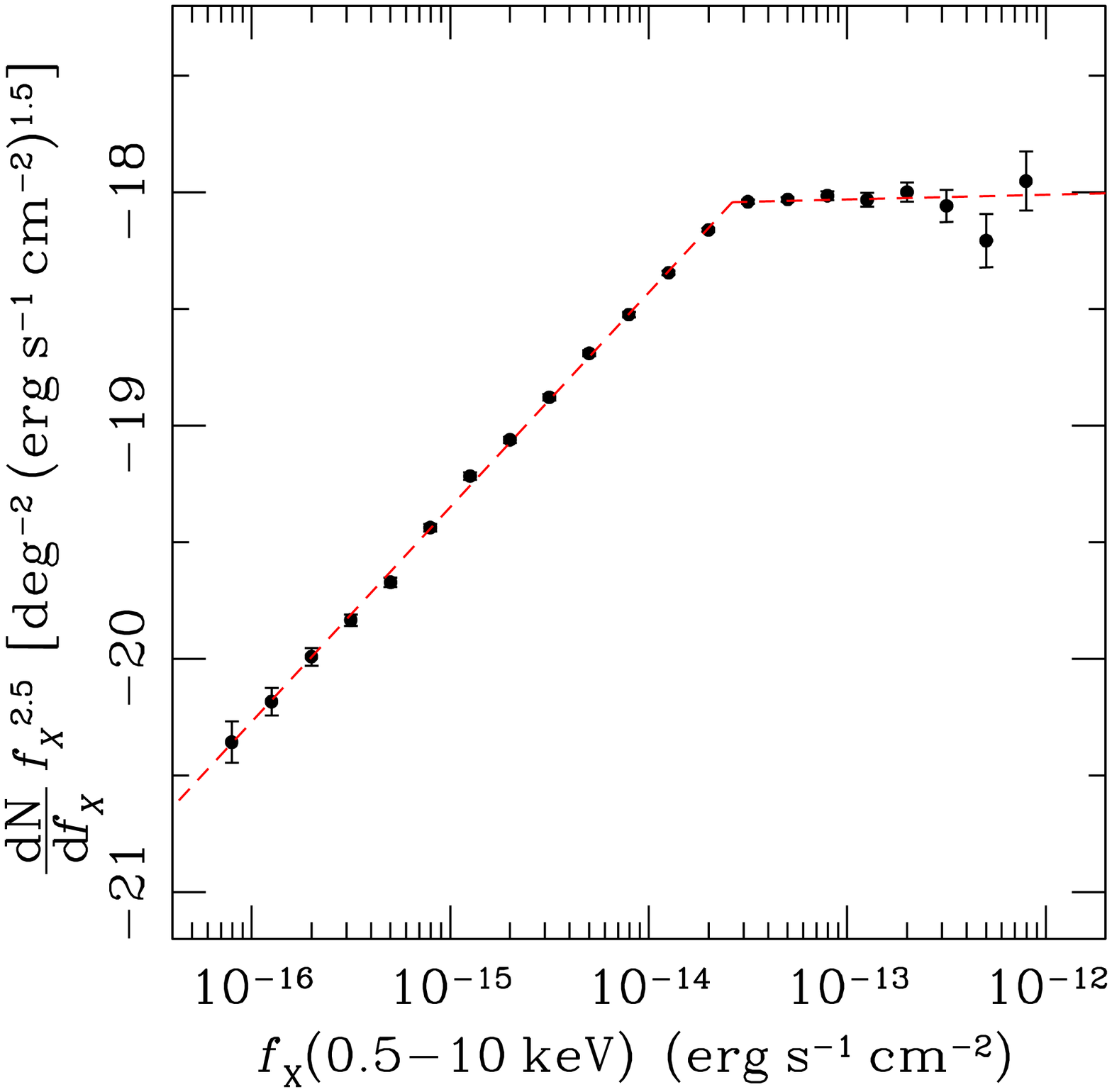}
}
\rotatebox{0}{  
  \includegraphics[height=0.9\columnwidth]{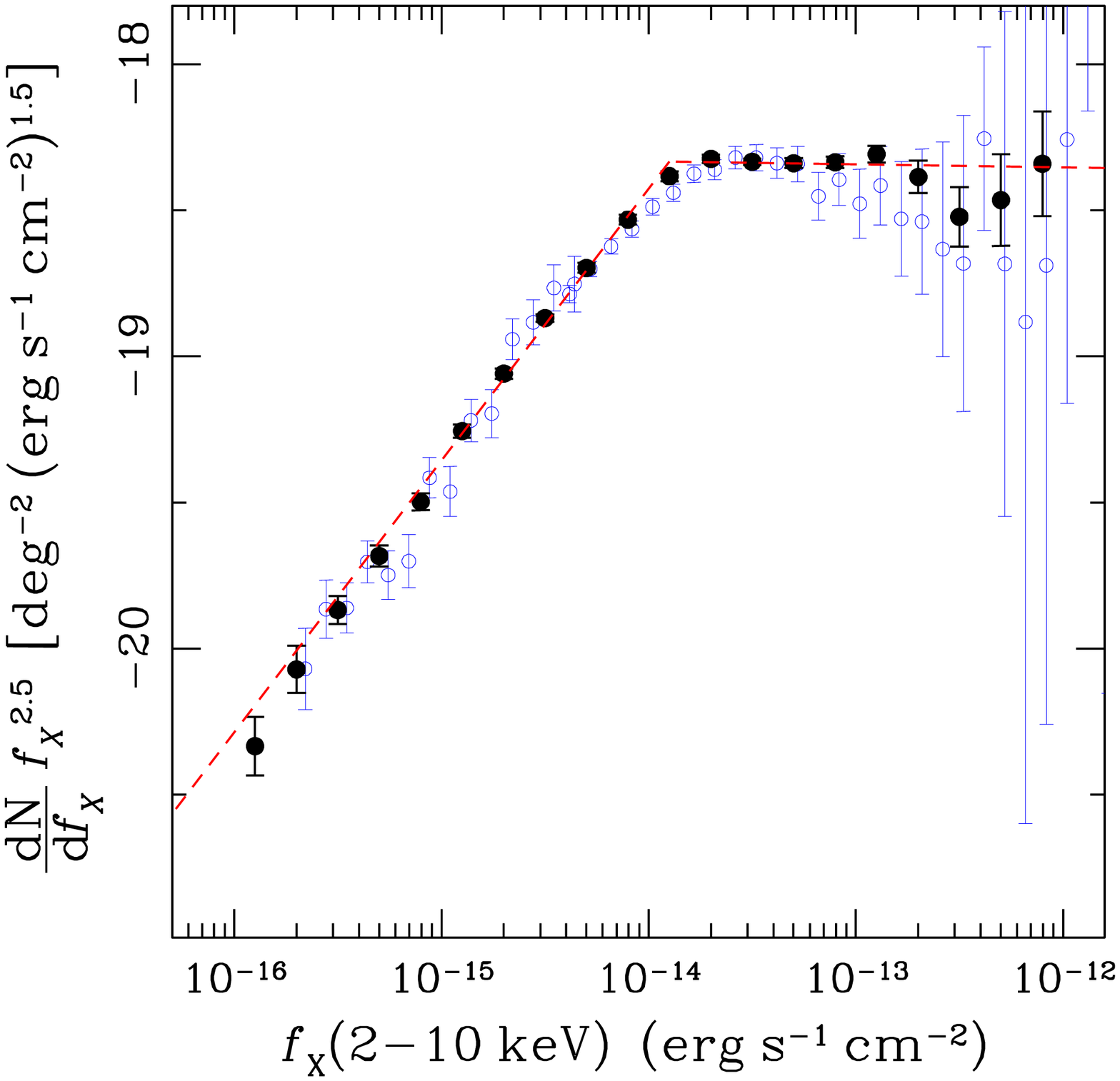}
  \includegraphics[height=0.9\columnwidth]{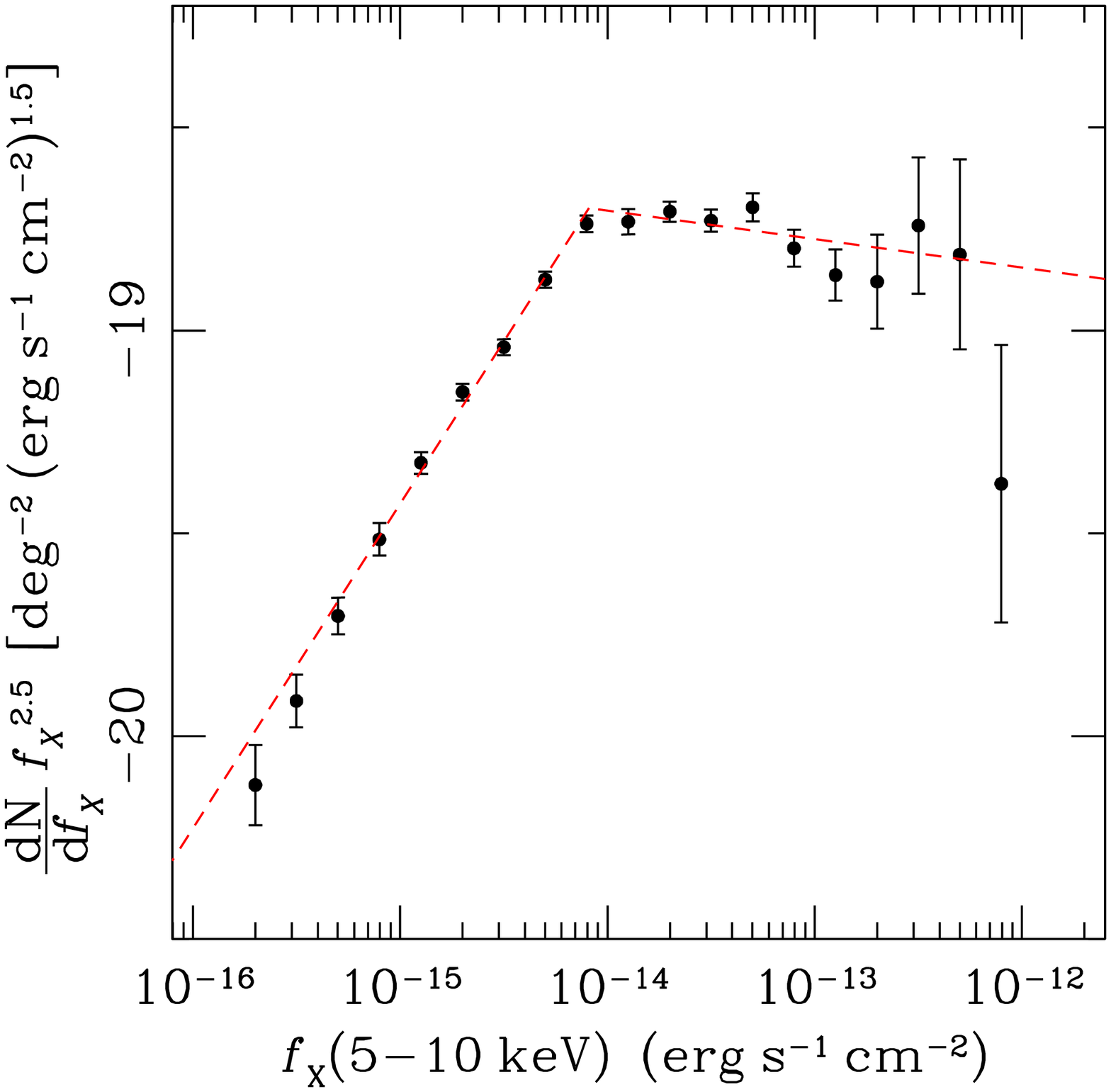}
}
\end{center}
\caption{  In   all  panels  the   (black)  filled  circles   are  the
differential number  counts normalised to  the Euclidean slope  in the
soft, total, hard and ultra-hard  bands for the combined {\it Chandra}
surveys listed  in Table \ref{tab_obs}.  The  error-bars are estimated
using bootstrap resampling as discussed  in the text. The (red) dashed
line in  each panel is  the maximum likelihood  fit to the  data.  The
(blue) open circles in the soft and the hard band count panels are the
differential number counts from the  ChaMP survey and the Chandra Deep
Fields  estimated by  Kim et  al.  (2007a).   For the  ultra-hard band
there  is  no  systematic  study  of  the  $dN/dS$  to  compare  with.
}\label{fig_dnds}
\end{figure*}

\begin{figure*}
\begin{center}
 \rotatebox{0}{
  \includegraphics[height=0.9\columnwidth]{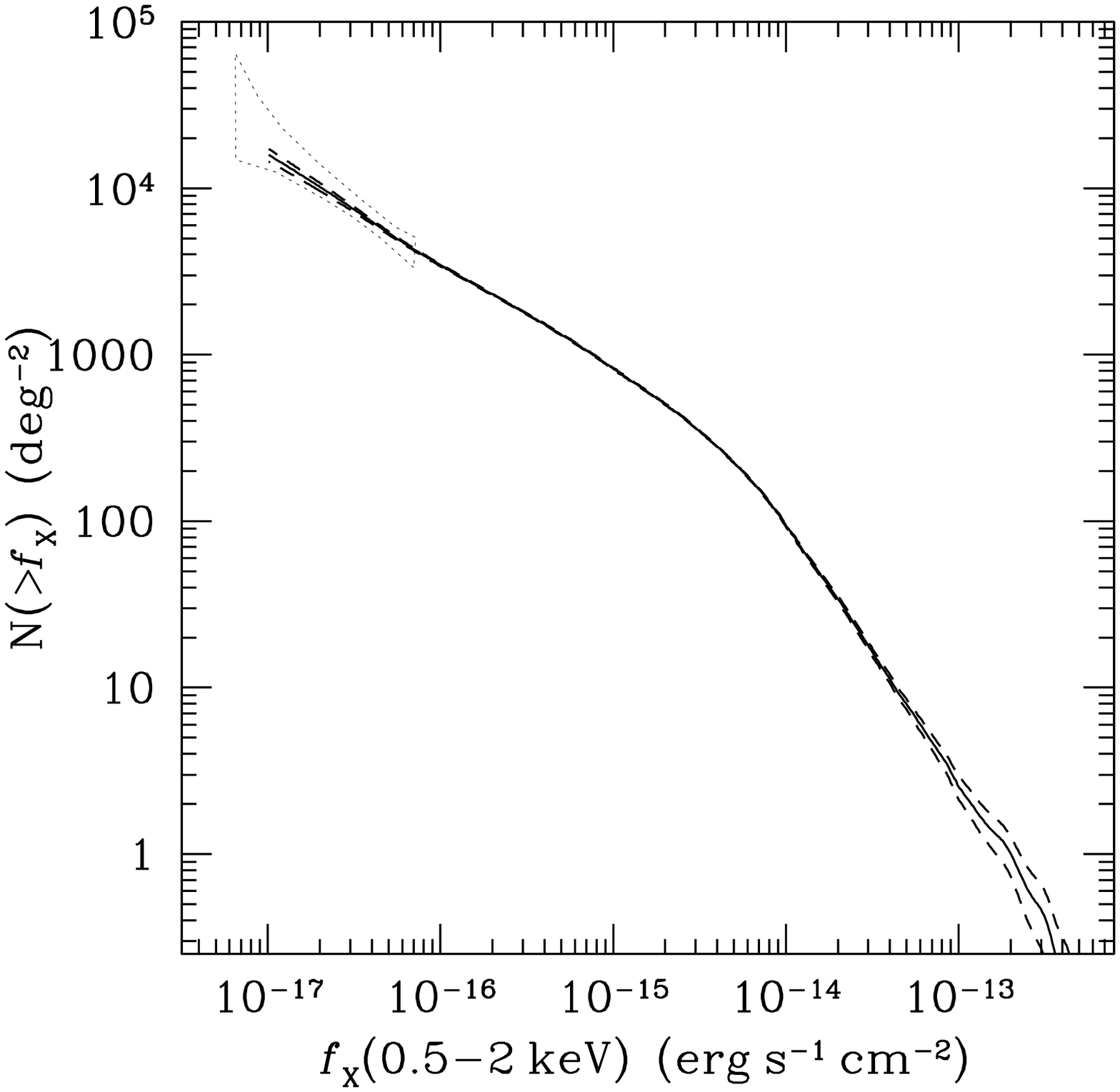}
  \includegraphics[height=0.9\columnwidth]{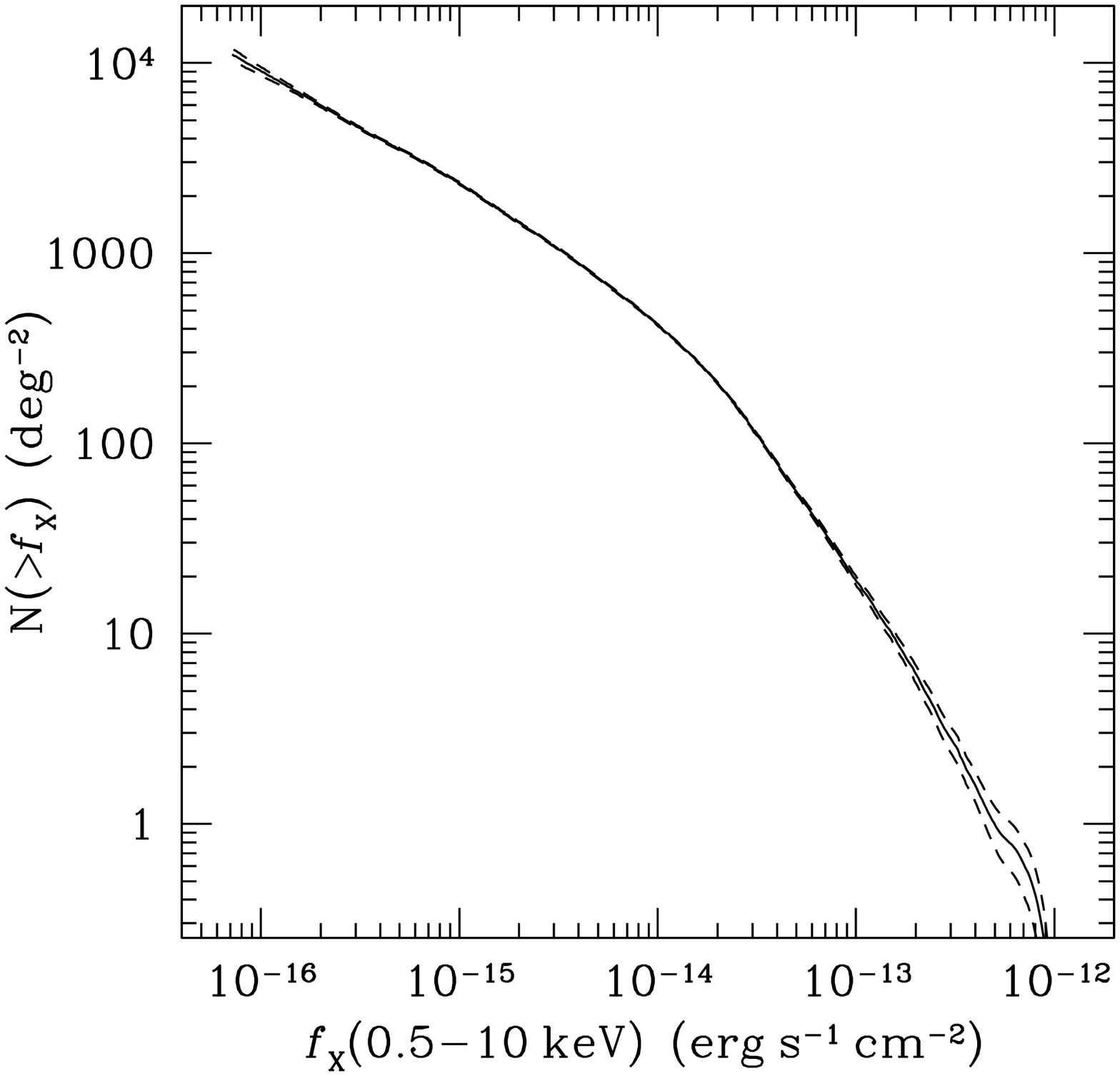}
}
\rotatebox{0}{  
  \includegraphics[height=0.9\columnwidth]{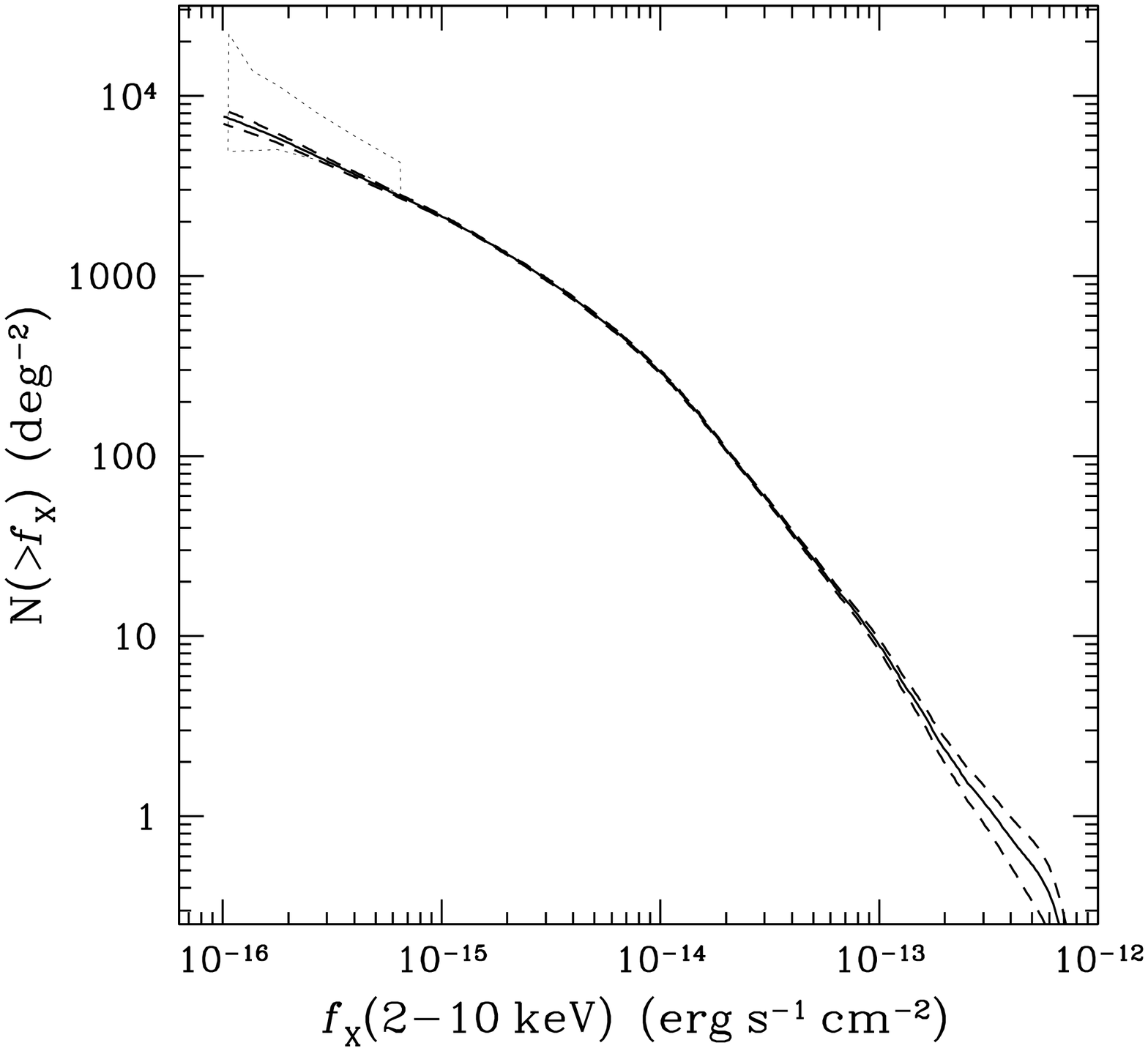}
  \includegraphics[height=0.9\columnwidth]{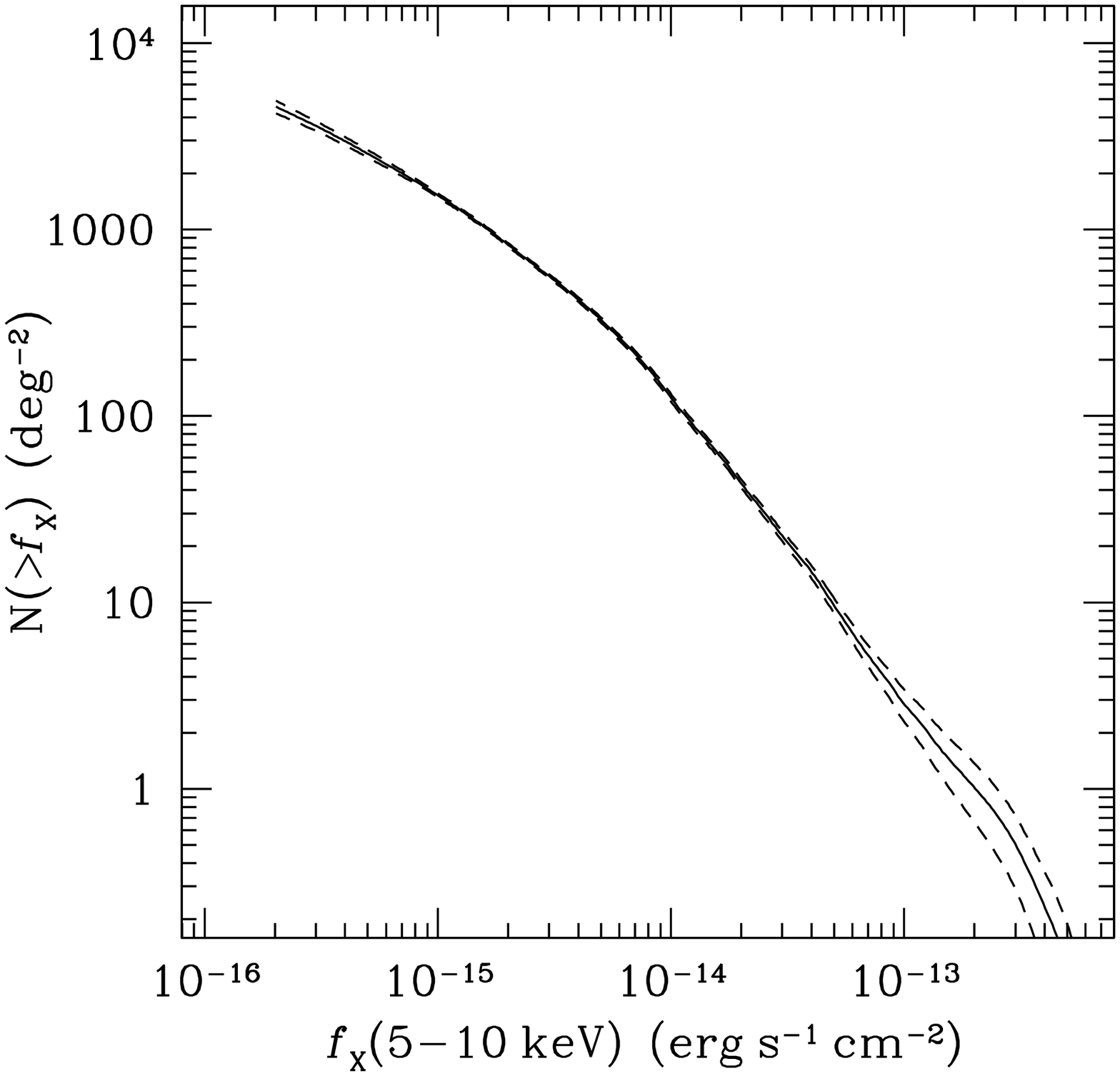}
}
\end{center}
\caption{
  Cumulative number counts in the soft, total, hard and
  ultra-hard bands for the combined {\it Chandra} surveys listed in
  Table \ref{tab_obs}. The continuous black line is the estimated
  number counts, the dashed lines correspond to the $1\sigma$ rms
  uncertainty estimated using bootstrap resampling of the data. For
  clarity the maximum likelihood fits to the data are not shown.  
  The 1\,Ms CDF-N fluctuation analysis results for the soft and hard
  bands (Miyaji \& Griffiths 2002) are shown with the dotted line 
  wedges. 
 }\label{fig_cum}
\end{figure*}

\begin{table} 
\caption{Stacking results}\label{tab_obs} 
\scriptsize
\begin{tabular}{l p{3.cm}  c  c c}
\hline 
Survey & Obs. IDs  & Exposure & Area           & Sources \\
       &           &  (ks)    & ($\rm deg^2$)  &         \\
  (1)  &  (2)      & (3)      & (4)            &  (5)    \\
\hline
CDF-N    & 580,   957,   966,   967 , 1671,   2232,  2233,  2234,
           2344,  2386,  2421,  2423,  3293,  3294,  3388-3391,  3408,
	   3409   & 2000 & 0.11 & 516 \\
CDF-S    & 581  441  582  1672 2405 2239 2312 2313 2406 2409 1431 & 900 & 0.06 & 270 \\
ECDF-S  & 5015-5022, 6164 & 250 & 0.25 & 592 \\
EGS    & 3305, 4357, 4365, 5841-5854, 6210-6223, 6366, 6391, 7169
         7180, 7181, 7187, 7188, 7236, 7237, 7238, 7239  & 200 & 0.63 &
         1325 
         \\ 
EN1    & 5855-5884 & 5 & 1.47 & 545 \\
XBOOTES    & 3596-3660, 4218-4272, 4277-4282 & 5 & 9.24 & 3056 \\
\hline
\end{tabular} 
\begin{list}{}{}
\item 
The columns are: (1): Survey  name; (2): {\it Chandra} observation IDs
used for each survey; (3) exposure time in ks. In the case of multiple
pointings this is  the mean exposure time; (4)  total surveyed area in
$\rm deg^2$. For  the ECDFS this includes the  overlap with the CDF-S;
(5) total number of sources  in each survey. The ECDFS sources include
those overlapping with the CDF-S.
\end{list}
\end{table}

\begin{table*}
\caption{X-ray number count best-fit parameters for the differential
  counts adopting the double power of equation \ref{eq_dnds}. The
  cumulative counts $N(>f_X)$ are obtained using equation
  \ref{eq_cum}}\label{tab_bestfit}  
\begin{center} 
\begin{tabular}{c c c   c c c}
\hline 
Band    & Sources & $\beta_1$  & $\beta_2$ & $\log f_{b}$ & $K$  \\
  (1)   &  (2)    & (3)        & (4)      &  (5)   & (6)  \\
\hline
Soft    & 4756    & $-1.58^{+0.02}_{-0.03}$ & $-2.50_{-0.05}^{+0.07}$ & $-0.04_{-0.05}^{+0.06}$ &  $1.51\pm0.03$ \\
Hard    & 2565    & $-1.56^{+0.04}_{-0.04}$ & $-2.52^{+0.07}_{-0.09}$ & $+0.09^{+0.08}_{-0.05}$ &  $3.79\pm0.08$ \\
Ultra-Hard & 1081  & $-1.70^{+0.08}_{-0.06}$ & $-2.57^{+0.07}_{-0.09}$ & $-0.09^{+0.06}_{-0.10}$ &  $2.36\pm0.08$ \\
Total   &  5561    & $-1.58^{+0.01}_{-0.02}$ & $-2.48_{-0.03}^{+0.06}$ & $+0.42_{-0.05}^{+0.07}$ &  $3.74\pm0.05$ \\

\hline

\end{tabular} 

\begin{list}{}{}
\item 

The columns are: (1): Spectral band; (2): Number of sources above the
$ 4\times10^{-6}$ detection threshold in a given energy band; (3) faint
end index of the double power-law; (4): bright end  
index of the double power law; (5): log of the break flux in units of 
$\rm 10^{-14} erg\,cm^{-2}\,s^{-1}$. (6): Normalisation in units $\rm
10^{16} deg^{-2}/(erg \, cm^{-2} \, s^{-1})$.

\end{list}

\end{center}
\end{table*}

\section{Application to real data}\label{results}

We apply the  methods developed in the previous  sections to real {\it
Chandra} data  to determine  the differential source  counts in  the 4
standard  X-ray spectral bands,  soft (0.5-2\,keV),  hard (2-10\,keV),
ultra-hard   (5-10\,keV)   and   total  (0.5-10\,keV).    We   combine
observations from  6 {\it Chandra} surveys, both  deep pencil-beam and
shallow wide-area.  Table  \ref{tab_obs} presents information on these
surveys,  which  include  the  Chandra  Deep  Field  North  and  South
(CDF-N/S), the Extended Chandra Deep Field South (ECDFS), the Extended
Groth  Strip (EGS),  the ELAIS-N1  (EN1) and  the XBOOTES  survey. The
combined  sample has  a  total  of 6295  unique  sources, detected  in
different  spectral bands,  over a  total area  of  $\rm 11.8\,deg^2$.
This is the  largest sample to date used for  the determination of the
X-ray number counts  and can only be compared with  the recent work of
Kim et  al. (2007a)  who combined observations  from the  CHAMP survey
(Kim et al. 2007b) with the Chandra Deep Fields.

The data from these surveys were  reduced and analysed in the same way
following methods  described by  Nandra et al.   (2005b) and  Laird et
al. (2008  in prep.).  Briefly, standard reduction  steps are followed
using the {\sc ciao} version 3.2 data analysis software.  Observations
corresponding  to the  same pointing  are merged  into a  single event
file.   The  ECDFS, EGS,  EN1  and  XBOOTES  surveys include  multiple
pointings, 4,  8, 30  and 126 respectively.   The CDF-S and  the ECDFS
observations  although  largely overlapping  are  treated as  separate
surveys.  This is to avoid  problems arising from merging regions with
significantly different  PSFs.  Images are constructed  in four energy
bands 0.5-7.0\,keV, 0.5-2.0\,keV,  2.0-7.0\,keV and 4.0-7.0\,keV.  The
X-ray catalogue for each spectral band is constructed using the source
detection method of section \ref{sec_det} adopting a Poisson detection
probability  threshold of  $4 \times  10^{-6}$.  The  total  number of
unique sources in each survey is shown in Table \ref{tab_obs}.  In the
case of duplicate sources in the overlap regions of adjacent pointings
in  the EGS,  ECDFS and  EN1 surveys  we keep  the detection  with the
smallest off-axis angle.   Also for the ECDFS we  exclude sources that
overlap with the deeper CDF-S survey.  The count rates in the 0.5-7.0,
0.5-2.0, 2.0-7.0 and 4.0-7.0\,keV bands are converted to fluxes in the
standard total,  soft, hard and ultra-hard bands  assuming a power-law
X-ray  spectrum  with   index  $\Gamma=1.4$  and  Galactic  absorption
appropriate for each  field. We note that this  assumption ignores the
hardening of the X-ray spectra  with decreasing flux (e.g. Mainieri et
al.  2002) or  the different  spectral properties  of  different X-ray
source populations (e.g. AGN vs normal galaxies).

We  apply  the methods  of  section  \ref{sec_sense}  to estimate  the
sensitivity curves of the surveys listed in Table \ref{tab_obs} taking
into account  overlapping regions (e.g.  ECDFS and  CDF-S).  These are
presented  in  Figure  \ref{fig_area}  for  the  hard  spectral  band,
2-10\,keV. We  follow the prescription of  section \ref{sec_counts} to
construct  and  to  fit  the  differential  counts  using  the  double
power-law equation  \ref{eq_dnds}.  The maximum  likelihood parameters
for   the   4   standard   spectral   bands  are   listed   in   Table
\ref{tab_bestfit}.   The results for  the differential  (normalised to
the  Euclidean slope)  and  cumulative number  counts  are plotted  in
Figures  \ref{fig_dnds}  and \ref{fig_cum}.   The  errorbars in  these
figures are  estimated using 100  bootstrap resamples of the  data. We
note that systematic uncertainties associated  with the use of a fixed
$\Gamma$ for the flux estimation or the EEF corrections are not taken
into account  in the calculation  of errors. With decreasing  flux the
area of the survey sensitive  to sources of that flux becomes smaller.
As a result  below a certain limit the observation  does not provide a
reliable census of the X-ray source population.  We choose to plot the
number counts  to the flux  limit corresponding to  1 per cent  of the
total  surveyed  area.  This  cutoff  applies  only  to the  graphical
representation  of  the number  counts  and  does  not affect  the  ML
calculation, where the decreasing  survey area with decreasing flux is
fully  accounted  for  in  the  calculation.  The  adopted  cutoff  is
typically  1.5-2 times  fainter  than  the standard  flux  limit of  a
particular observation (see Figure 2).   As a result the number counts
derived  here extend  to  fluxes  that are  1.5-2  times fainter  than
previous   determinations.     This   is   demonstrated    in   Figure
\ref{fig_dnds} for  the 0.5-2 and 2-10\,keV bands,  for which accurate
estimates of the differential  counts are available in the literature.
Our work also has the advantage  that all the fields used to determine
the number  counts have  been analysed in  a homogeneous way  and that
{\it all} the detected sources, even  those close to the flux limit of
the surveys in Table \ref{tab_obs}, have been used in the calculation.
This is  because our approach  for determining the  $\log N -  \log S$
correctly  accounts for  the completeness  and flux  bias corrections,
particularly for sources with few photons close to the detection limit
of a given survey.

The best-fit parameters  for the 0.5-2, 2-10 and  0.5-10\,keV bands in
Table \ref{tab_obs} are in good agreement with recent estimates by Kim
et  al.   (2007a), who  also  combined  deep  pencil-beam surveys  with
shallow observations over a wide  area with {\it Chandra} to determine
the  number counts  in  the  0.5-2, 2-8  and  0.5-8\,keV bands.   Also
presented here  for the  first time, is  the $\log  N -\log S$  in the
5-10\,keV band over  4\,dex in flux.  This wide  flux range allows the
determination  of  the flux  of  the  break  in the  double  power-law
representation of the 5-10\,keV number counts.

\begin{figure}
\begin{center}
 \rotatebox{0}{\includegraphics[height=0.9\columnwidth]{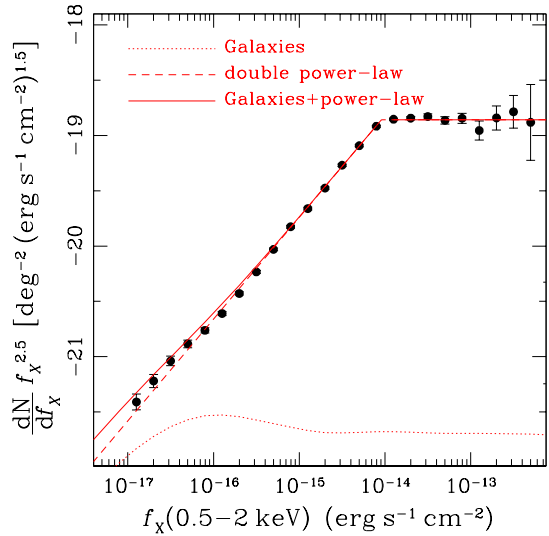}}
\end{center}
\caption{ Soft-band differential source  counts in comparison with the
$dN/df_X$  for star-forming  galaxies (Georgakakis  et al.  2007). The
surface density  of these  sources can account  for the  excess counts
above the  broken double power-law  expectation at fluxes  below about
$\rm   6\times    10^{-17}   \,    erg   \,   s^{-1}    \,   cm^{-2}$.
}\label{fig_dnds_gals}
\end{figure}

In Figure \ref{fig_dnds} the soft band source counts at fluxes fainter
than  about $\rm  6\times 10^{-17}\,  erg  \, s^{-1}  \, cm^{-2}$  lie
systematically above the best-fit double power-law. The emergence of a
population of normal star-forming galaxies at faint fluxes can explain
this  excess.  This  is  demonstrated  in  Figure  \ref{fig_dnds_gals}
plotting the expected star-forming  galaxy number counts, estimated by
integrating  the X-ray  luminosity function  of these  systems  at low
redshift (Georgakakis et al.  2006) assuming pure luminosity evolution
of the  form $\propto (1+z)^{2.4}$  (e.g.  Georgakakis et  al.  2007).

Contrary  to the  soft-band, the  2-10 and  5-10\,keV counts  at faint
fluxes in Figure \ref{fig_dnds} show tenative evidence, significant at
$\approx2\sigma$ level, for a flattening of the faint-end slope at the 
limit of the  deepest X-ray survey of the  sample, the CDF-North.  Kim
et al.  (2007a) have shown that  the $\log N - \log S$ distribution of
sources depends on their hardness  ratio, HR.  Sources with $\rm HR>0$
have steep differential  counts that do not show  evidence for a break
in  the slope  at the  flux limit  of the  ChaMP survey.  In contrast,
sources with $\rm HR<0$, show the characteristic break in their number
count distribution.  It is suggested that sources with $\rm HR>0$ lie,
on average, at redshifts lower than the population with $\rm HR<0$ and
as a  result they  do not show  the cosmological  evolutionary effects
that cause the break  in the $\log N - \log S$  (Harrison et al. 2003;
Kim et al.  2007a).  The change  in slope at the faint-end of the 2-10
and 5-10\,keV counts is likely related to the relative contribution of 
sources with  $\rm HR\lessgtr0$. 

\begin{figure*}
\begin{center}
 \rotatebox{0}{
  \includegraphics[height=0.9\columnwidth]{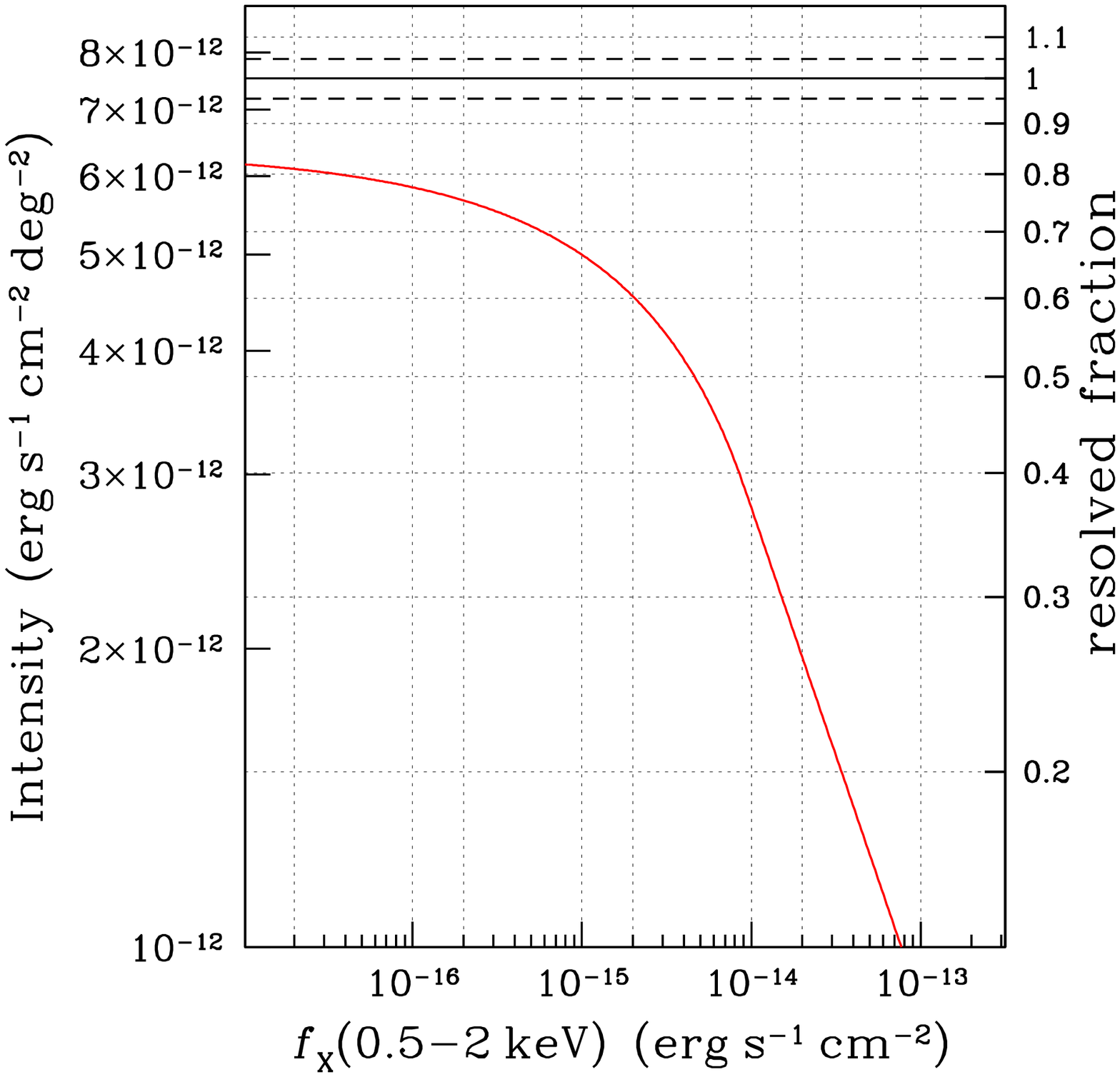}
  \includegraphics[height=0.9\columnwidth]{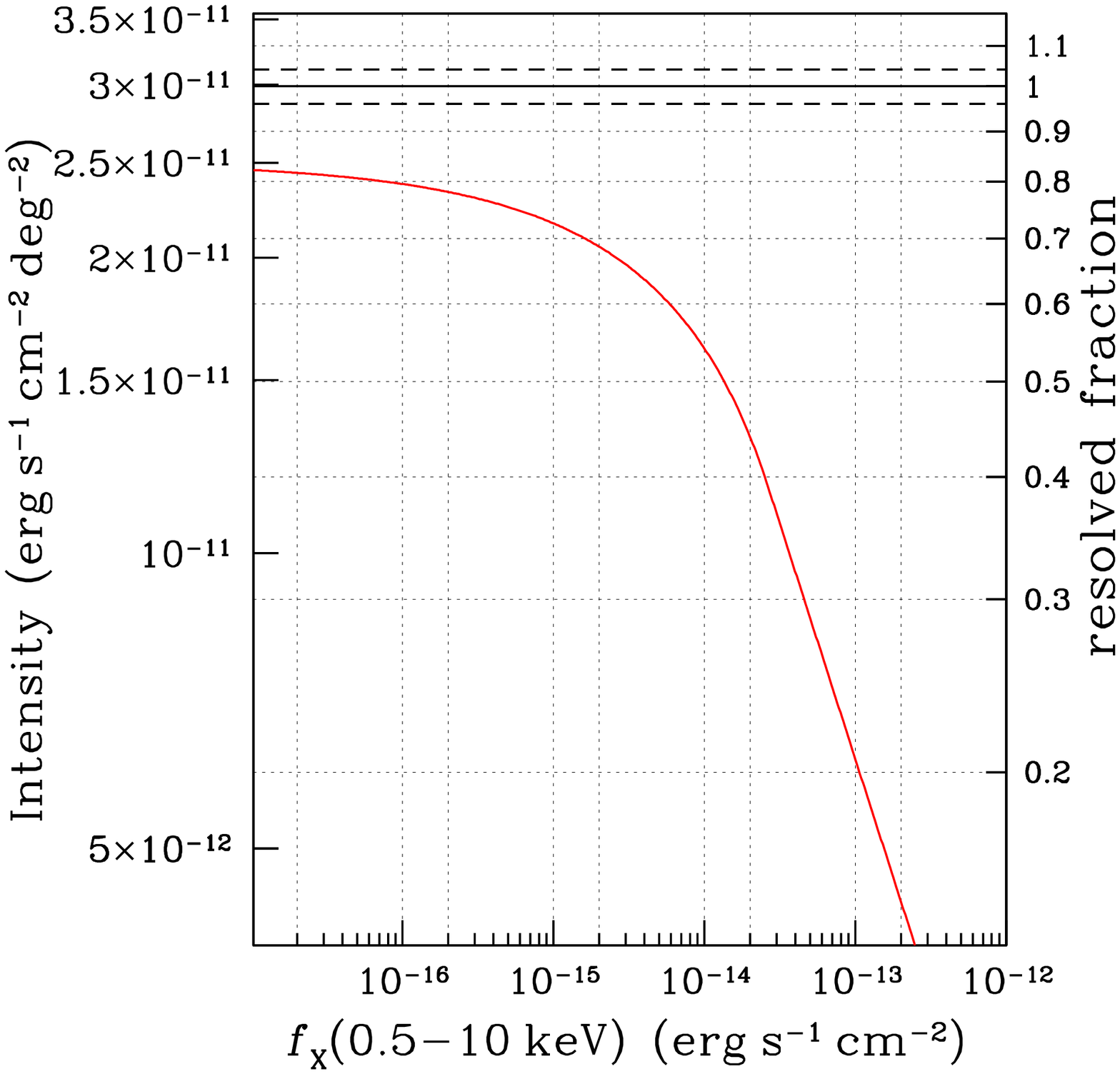}
}
\rotatebox{0}{  
  \includegraphics[height=0.9\columnwidth]{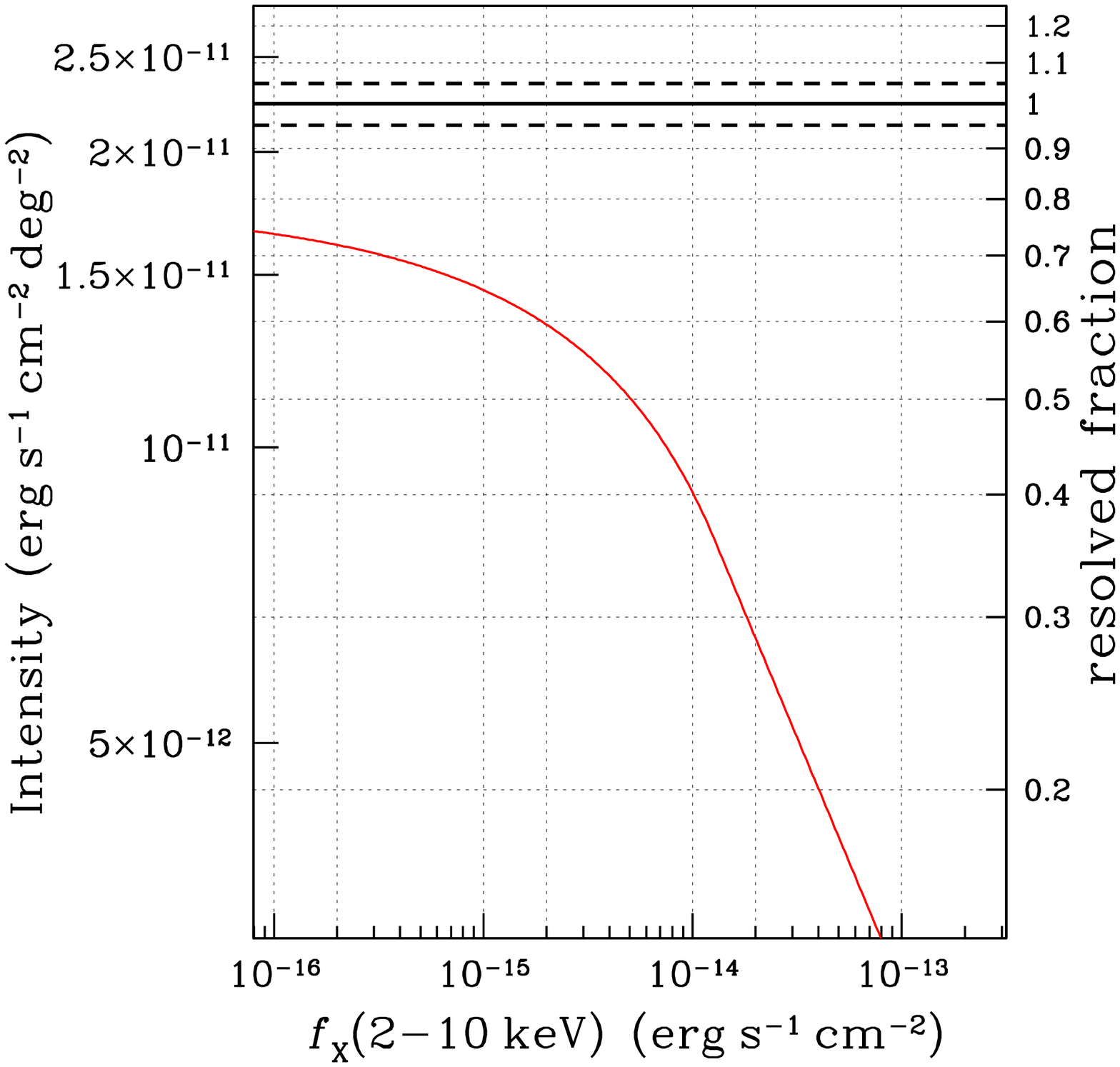}
  \includegraphics[height=0.9\columnwidth]{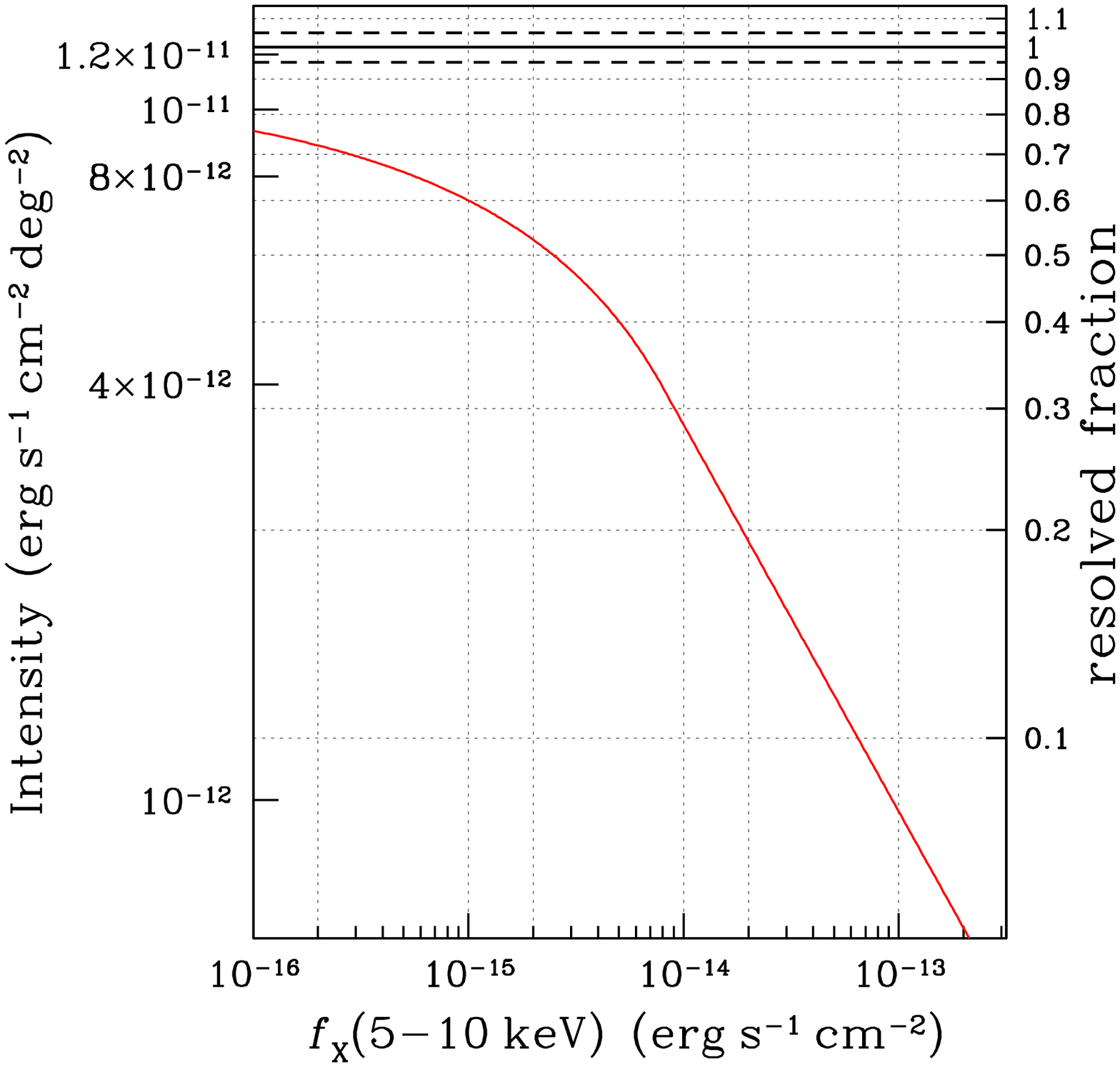}
}
\end{center}
\caption{
  Integrated intensity and contribution to the diffuse
  X-ray background of the point sources detected in different spectral
  bands. The (red) curves are estimated from the  maximum likelihood
  fits to the differential number counts. The level of the X-ray
  background is shown with the horizontal continuous line. 
  The dashed lines correspond to the statistical $1\sigma$ rms
  uncertainty to this  value.    
 }\label{fig_xrb}
\end{figure*}

Figure \ref{fig_xrb}  plots the contribution  of the point  sources to
the diffuse  X-ray background (XRB) in different  spectral bands. This
is estimated  by integrating the  double power-law relation  using the
best-fit parameters listed in  Table \ref{tab_bestfit}.  For the level
of XRB  we adopt the  average flux densities of  $(7.52\pm0.35) \times
10^{-12}  \rm  \,erg  \,  s^{-1}  \,  cm^{-2}  \,  deg^{-2}$  for  the
0.5-2\,keV  band from  Moretti et  al.   (2003) and  $(2.24 \pm  0.11)
\times 10^{-11}  \rm \,erg \, s^{-1}  \, cm^{-2} \,  deg^{-2}$ for the
2-10\,keV  range  from De  Luca  \&  Molendi  (2004).  For  the  total
spectral band,  0.5-10\,keV, we add  the values above and  estimate an
XRB flux  density of $(2.99\pm0.12)\times10^{-11} \rm  \,erg \, s^{-1}
\, cm^{-2} \,  deg^{-2}$. In the case of the  5-10\,keV band we assume
$\Gamma=1.41$ and the 2-10\,keV XRB flux density determined by De Luca
\& Molendi  (2004) to estimate  $\rm (1.23\pm0.06)\times10^{-11} \,erg
\, s^{-1} \,  cm^{-2} \, deg^{-2}$.  The fraction  of the XRB resolved
in  point sources  in  different energy  bands  to the  limits of  the
deepest survey  in the  sample are listed  in Table  \ref{tab_xrb}. In
addition to the statistical uncertainty  listed in that table there is
a systematic error of about  10-20 per cent related to the uncertainty
in the  determination of the  absolute normalisation of the  XRB (e.g.
Revnivtsev et al.   2003, 2005).  In Table \ref{tab_xrb}  the 2-10 and
5-10\,keV  bands resolve similar  fractions of  the XRB,  $74\pm4$ and
$72\pm4$ per cent  respectively.  This is consistent with  the work of
Worsley et al. (2005). In that  study the fraction of the XRB resolved
into  point  sources  is  nearly  constant  at  $2-6$\,keV  and  drops
substantially  at energies  $>6$\,keV  to  about 50  per  cent in  the
$8-12$\,keV band. 

The resolved XRB fractions in the 2-10 and 5-10\,keV bands can be used
to  place constraints  on the  spectral shape  of the  unresolved XRB.
First, the difference in the sensitivity in the two energy bands needs
to be accounted for by matching their flux limits.  The 5-10\,keV flux
limit  of  $2\times  10^{-16}  \rm   \,  erg  \,  s^{-1}  \,  cm^{-2}$
corresponds to  $f_X(2-10)=3.6\times 10^{-16} \rm \, erg  \, s^{-1} \,
cm^{-2}$ for  $\Gamma=1.4$, i.e.   the mean spectrum  of the  XRB.  At
this  flux $70\pm4$  per cent  of  the XRB  in the  2-10\,keV band  is
resolved into  point sources.  We then  assume that there  is a single
population that is  responsible for the unresolved part  of the XRB in
both  the $2-10$  ($30\pm4$ per  cent) and  $5-10$\,keV  ($28\pm4$ per
cent)  bands.  There has  been  speculation  recently  on whether  the
unresolved background  at hard energies  is produced by  Compton thick
AGN (Worsley et al. 2005; Gilli  et al.  2007).  In order to test this
possibility we adopt  for the X-ray spectral shape  of this population
the  Compton reflection  models of  Magdziarz \&  Zdziarski  (1995) as
implemented in  the {\sc pexrav} spectral energy  distribution of {\sc
xspec}.   We assume  a mean  redshift  $z\approx1$, a  solid angle  of
$2\pi$, solar  abundance for all  elements and an  average inclination
relative  to the  line of  sight $\cos  i=0.45$.  Only  the reflection
component  was used, i.e.  no direct  radiation. We  find that  if the
entire 30 per cent of  the unresolved background in the 2-10\,keV band
(to the  limit $f_X(2-10)=3.6\times 10^{-16}  \rm \, erg \,  s^{-1} \,
cm^{-2}$)  is produced  by such  sources, then  their  contribution to
5-10\,keV XRB  is 39 per  cent, which exceeds the  unresolved fraction
($28\pm4$ per cent) at the $2.8\sigma$ level. Moving the mean redshift
of  the  Compton  thick  AGN  population to  $z\approx2$  reduces  the
significance of  the excess to  $2.2\sigma$.  We conclude  that either
some of  the sources  in the 5-10\,keV  selected sample lie  below the
flux  limit $f_X(2-10)=3.6\times  10^{-16}  \rm \,  erg  \, s^{-1}  \,
cm^{-2}$ or the  unresolved XRB fraction in the  2-10\,keV band cannot
be entirely due to a single population of Compton thick AGN.

\begin{table} 
\caption{Resolved X-ray Background fractions}\label{tab_xrb} 
\begin{tabular}{l c c}
\hline 
Band    & limit & fraction  \\
  (1)   &  (2)    & (3)     \\
\hline
Soft    & $1\times10^{-17}$    & $82\pm4$\\
Hard    & $1\times10^{-16}$    & $74\pm4$\\
Ultra-Hard &  $2\times10^{-16}$  & $72\pm4$ \\
Full   &  $7\times10^{-17}$    &  $81\pm3$ \\
\hline
\end{tabular} 
\begin{list}{}{}
\item 
The columns are: (1): X-ray energy band; (2): flux limit in the
corresponding spectral band that the resolved XRB fraction is
estimated. The units are $\rm erg\,s^{-1}\,cm^{-2}$. At these flux
limits the area curve of the deepest survey in the sample, the CDF-N,
drops to about 1 per cent of its maximum value; (3) Per cent fraction of the
XRB resolved into point sources.  
\end{list}
\end{table}

\section{Conclusions}

A new  method is  presented for determining  the sensitivity  of X-ray
imaging  observations, which accurately  estimates the  probability of
detecting   a  source  with   a  given   X-ray  flux   accounting  for
observational effects, such as  vignetting, flux estimation biases and
the fraction of  spurious sources expected in any  source catalogue. A
major advantage  of the proposed method  is that it  is analytical and
therefore does  not require a  large number of cumbersome  ray tracing
simulations to quantify the effects  above.  We demonstrate how to use
the  sensitivity  maps  determined  by  our new  method  in  order  to
accurately  estimate the  number  counts in  different X-ray  spectral
bands, using all the detected sources, even those with few counts, for
which  the completeness  and flux  bias corrections  are  large.  This
method is applied to real {\it Chandra} data. The sample includes both
deep pencil-beam  and shallow wide-area surveys covering  a total area
of about $\rm  11.8\,deg^2$ and includes over 6000  unique sources. We
present, for  the first time, the  X-ray counts in  the 5-10\,keV band
over a  wide range  of X-ray fluxes  (4\,dex) and determine  the break
flux  in  this band.  We  also  find evidence  for  an  upturn in  the
0.5-2\,keV differential  counts below  about $\rm 6\times  10^{-17} \,
erg \,  s^{-1} \, cm^{-2}$, which  we attribute to the  emergence of a
population  of star-forming galaxies  at faint  fluxes.  Based  on the
fraction of the XRB resolved in  the 2-10 and 5-10\,keV bands we argue
that a single population of Compton thick AGN cannot by itself produce
the entire unresolved X-ray background in the 2-10\,keV energy range.

\section{Acknowledgements}

We thank the anonynous referee for providing constructive comments and
suggestions that improved this paper.  This work has been supported by
funding from the Marie-Curie Fellowship grant MEIF-CT-2005-025108 (AG)
and the STFC (ESL). The data products used in this paper are available
for  download at
http://astro.ic.ac.uk/research/xray/chandrasurveys.shtml

\end{document}